\documentclass[superscriptaddress, noshowpacs, noshowkeys, twocolumn, pra]{revtex4}
\usepackage{amsmath,amssymb,amsfonts,latexsym}
\usepackage{color}

\usepackage[colorlinks,bookmarks=false,citecolor=blue,linkcolor=red,urlcolor=blue]{hyperref}

\usepackage[T1]{fontenc}
\usepackage[utf8]{inputenc}
\usepackage[english]{babel}
\usepackage{amsmath,mathrsfs,
color,mathdots, bbold,tikz, xspace,version}
\usepackage{epsfig}
\usepackage{graphicx}
\usepackage{epstopdf}

\newcommand{\ket}[1]{\ensuremath{|#1\rangle}\xspace}
\newcommand{\bra}[1]{\ensuremath{\langle #1|}\xspace}

\newcommand\be{\begin{equation}}
\newcommand\bea{\begin{eqnarray}}
\newcommand\ee{\end{equation}}
\newcommand\eea{\end{eqnarray}}

\begin{document}

\title{Decoherence of Bell states by local interactions with a suddenly quenched spin environment }

\author{Pierre Wendenbaum}
\affiliation{Institut Jean Lamour, dpt P2M, Groupe de Physique Statistique, Universit\'e de Lorraine-CNRS, B.P. 70239, F-54506 Vand\oe{}uvre-l\`es-Nancy Cedex, France}
\affiliation{Theoretische Physik, Universit\"at des Saarlandes, D-66123 
Saarbr\"ucken, Germany}
\author{Bruno G. Taketani}
\affiliation{Theoretische Physik, Universit\"at des Saarlandes, D-66123 
Saarbr\"ucken, Germany}
\author{Dragi Karevski}
\affiliation{Institut Jean Lamour, dpt P2M, Groupe de Physique Statistique, Universit\'e de Lorraine-CNRS, B.P. 70239, F-54506 Vand\oe{}uvre-l\`es-Nancy Cedex, France}

\date{\today}

\begin{abstract}
We study the dynamics of disentanglement of two qubits initially prepared in a Bell state and coupled at different sites to an  Ising transverse field spin chain (ITF) playing the role of a dynamic spin environment. The initial state of the whole system is prepared into a tensor product state $\rho_{Bell}\otimes \rho_{chain}$ where the state of the chain is taken to be given by the ground state $|G(\lambda_i)\rangle$ of the ITF Hamiltonian $H_E(\lambda_i)$ with an initial field $\lambda_i$. 
At time $t=0^+$,  the strength of the transverse field is suddenly quenched to a new value $\lambda_f$ and the whole system (chain $+$ qubits) undergoes a unitary dynamics generated by the total Hamiltonian $H_{Tot}=H_E(\lambda_f) + H_I$ where $H_I$ describes a local interaction between the qubits and the spin chain. The resulting dynamics leads to a disentanglement of the qubits, which is described through the Wooter's Concurrence, due to there interaction with the non-equilibrium environment.  The concurrence is related to the Loschmidt echo which in turn is expressed in terms of the time-dependent covariance matrix associated to the ITF. This permits a precise numerical and analytical analysis of the disentanglement dynamics of the qubits as a function of their distance, bath properties and quench amplitude. In particular we emphasize the special role played by a critical initial environment.\end{abstract}


\maketitle

\section{Introduction}
Entanglement is one of the most intriguing features of nature 
\cite{hor09} predicted by quantum mechanics. Since
the pioneering discussion by Einstein, Podolsky and Rosen in there 
celebrated paper
\cite{Ein35}, the dramatic consequences of quantum entanglement have 
been extensively studied on both theoretical and experimental sides (see 
\cite{Ber06} for an historical review).
If these initial studies were first orientated to a better understanding 
of the foundations of quantum mechanics, more recent investigations on 
entanglement phenomena focused on potential technological applications 
such as quantum computing \cite{Nie00} and quantum simulation 
\citep{Geo14}.

However, entanglement is generally very sensitive to decoherence 
generated by the unavoidable interactions with the system's environment 
\cite{Paz01,Zur02, Sch07}, responsible for the loss of the typical 
quantum features one wishes to exploit. It is consequently of primary 
importance to understand these decoherence processes in order to 
suppress or possibly exploit it. For example, in order to limit the 
decoherence process, dynamical control consisting in pulses applied to 
the system has been proposed in \cite{Vio98,Ros08}. Engineered 
non-equilibrium dynamics have also been suggested to create entangled 
steady-states \cite{Fog13,Tak14}  and to assist precision measurements \cite{PhysRevLett.106.140502}. 
From a different perspective, typical quantum 
information tools such entanglement have been applied in many-body 
systems to identify signatures of quantum phase transitions 
\cite{Cin07}  and to characterize the 
ground state close to a critical point \cite{ost02}.

Aiming at a better understanding of decoherence, a number of models 
investigated the dynamics of a small system interacting with a given 
typical environment. Among them one may mention the central spin model, 
where the system made of one or two spins is simultaneously coupled to 
many interacting spins 
\cite{Cuc05,Qua06,cuc07,Yua07,Dam11,Muk12,Sha12,Nag12,Far13}. Particular 
focus has been set on critical spin environments which were shown to 
lead to enhanced decoherence \cite{Qua06} and to universal properties 
\cite{cuc07}. Cormick and Paz \cite{Cor08} went beyond the standard 
central spin system and studied the dependence of decoherence on the 
spatial separation of two qubits, initially prepared in a Bell state, 
when they interact locally with an extended equilibrium environment 
modeled by a quantum spin $1/2$ chain in a transverse field.  They found 
in particular that in the strong coupling limit decoherence typically increases with the qubits 
separation distance and finally saturates when the qubits separation is 
over a threshold distance related to the spin chain correlation length.

In this work, we extend Cormick and Paz work \cite{Cor08}  by considering an environment 
which is set out of equilibrium by a sudden change of a global 
environment coupling constant, the so called global quantum quench 
\cite{Cal07,Pol11}. Quantum quench protocols have received these recent 
years much attention as for example in the context of the quantum 
version of fluctuation theorem \cite{Dor08}, the relaxation properties 
toward a local canonical ensemble or a generalized version of Gibbs 
ensemble depending on the integrability of the system, see \cite{Pol11} 
for a review.
Many of these investigations focused not only on steady properties but 
also on dynamical aspects like front propagation of an initial density 
inhomogeneity \cite{Kar02,Hun04,Pla05,Pla07} or the expansion of a cloud 
of particles after the more or less sudden release of a trap 
\cite{Col10,Col11,Col12,Wen13}.
Our main goal here is to investigate how the quench, that is how the 
relaxation of the environment toward a local steady state \cite{Pol11},  
influences, with respect to the equilibrium case treated in 
\cite{Cor08}, the disentanglement of the two distant qubits initially 
prepared in a Bell state.

The paper is organized as follows: In section II we present the model describing two qubits coupled to an Ising Chain in a Transverse Filed (ITF). In section III the dynamics is diagonalized through the Jordan Wigner representation of the ITF and an explicit relation is given for the Loschmidt echo through the time evolution of the two-point correlation functions of the ITF. Section IV is devoted to the quench behavior of the disentanglement of the qubits studied numerically and analytically. 
Finally in section V we draw our conclusions.

\section{The model and the entanglement measure}
\subsection{Two qubits coupled to an Ising chain}
\begin{figure}[h!]
\begin{center}
\includegraphics[scale=0.5]{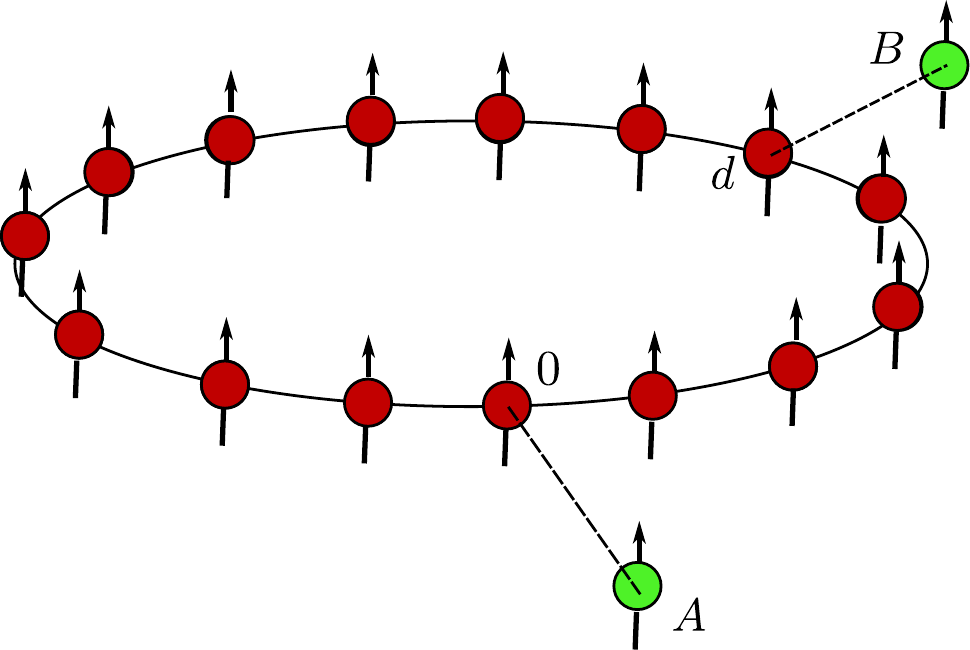}
\caption{(Color online) Two defect spins  (qubits) $A$ and $B$ are locally coupled to a spin chain.}
\label{fig1}
\end{center}
\end{figure}

We consider in the following two non-interacting qubits coupled locally to an Ising quantum chain with $N$ spins (see figure \ref{fig1}).  
The total Hamiltonian (qubits + chain) governing the dynamics of the whole system is given by
\begin{equation}
H_{Tot}=H_E+H_{I}\; ,
\end{equation}
where $H_E$ is the Ising chain (environment) Hamiltonian
\be
H_E(\lambda)=-J\sum_{j=0}^{N-1}\sigma_j^x\sigma_{j+1}^x
-\lambda\sum_{j=0}^{N-1}\sigma_j^z,
\label{IsingH}
\ee
where the $\sigma$'s are the usual Pauli matrices. The nearest neighbor coupling $J$ is taken to be positive and $\lambda$ is a transverse field. We work with periodic boundary conditions, i.e $\sigma^i_{N}=\sigma^i_{0}$ with $i=x,y,z$. The interaction Hamiltonian describing the coupling of the qubits, labeled  $A$ and $B$, at different sites of the chain separated by a distance $d$ is given by
\begin{equation}
H_I=-\varepsilon \big(\ket{\uparrow}\bra{\uparrow}_A\otimes \sigma_0^z+\ket{\uparrow}\bra{\uparrow}_B\otimes \sigma_d^z\big)\; ,
\end{equation}
where $\ket{\uparrow}_{A,B}$ is an eigenstate of $\sigma^z_{A,B}$ satisfying $\sigma^z_{A,B}\ket{\uparrow}_{A,B}=\ket{\uparrow}_{A,B}$ and $\varepsilon>0$ sets the intensity of that interaction. 

The two qubits are assumed to be initially in the maximally entangled Bell state 
$\ket{\phi}_{AB}=\frac{1}{\sqrt{2}}(\ket{\uparrow \uparrow}+\ket{\downarrow \downarrow})$
and uncorrelated to the bath, such that the initial state of the total system is a tensor state $\ket{\psi(0)}=\ket{\phi}_{AB}\otimes \ket{G(\lambda_i)}_E$, with $\ket{G(\lambda_i)}_E$ the ground state of the initial bath Hamiltonian $H_E(\lambda_i)$. 

At time $t=0^+$ the transverse field of the Ising chain is suddenly quenched to a new value $\lambda_f$ forcing the system to evolve in a non-equilibrium regime. 
Due to the structure of the interaction Hamiltonian and the initial state, 
the total dynamics splits into two different channels, each governed by a specific Hamiltonian, namely $H_{\downarrow \downarrow}(\lambda_f)=H_E(\lambda_f)$ if the two qubits are in the state $\ket{\downarrow \downarrow}$ and $H_{\uparrow \uparrow}(\lambda_f)=H_E(\lambda_f)-\varepsilon(\sigma_0^z+\sigma_d^z)$ if they are in the state $\ket{\uparrow \uparrow}$. 
Notice here that $H_{\uparrow \uparrow}$ has exactly the same structure as $H_E$, the only difference being that the transverse fields acting at sites $0$ and $d$ are changed to the value $\lambda+\varepsilon$ instead of $\lambda$. 
Consequently, the time evolution of the initial state  $\ket{\psi(0)}=\ket{\phi}_{AB}\otimes \ket{G(\lambda_i)}$ is given by
\begin{equation}
\ket{\psi(t)}=\frac{1}{\sqrt{2}}\big[\ket{\uparrow \uparrow}\otimes \ket{\varphi_{\uparrow \uparrow}(t)}_E+\ket{\downarrow \downarrow}\otimes \ket{\varphi_{\downarrow \downarrow}(t)}_E\big] \; ,
\end{equation}
with the  evolved states  
\be
\ket{\varphi_{\alpha}(t)}_E=e^{-iH_{\alpha}(\lambda_f)t}\ket{G(\lambda_i)}_E
\ee
where $\alpha={\uparrow \uparrow},{\downarrow \downarrow}$.

The reduced density matrix of the qubits, $\rho_s(t)=Tr_E\{\ket{\psi(t)}\bra{\psi(t)}\}$, is given in the computational base $\{\ket{\uparrow \uparrow}, \ket{\uparrow \downarrow}, \ket{\downarrow \uparrow}, \ket{\downarrow \downarrow}\}$ by
\begin{equation}
\rho_s(t)=\frac{1}{2}
\begin{pmatrix}
1  & 0 & 0 & D_{\uparrow \uparrow,\downarrow \downarrow}(t)\\
\\
0& 0 & 0 &0 \\
\\
0& 0 & 0 &0 \\
\\
D^*_{\uparrow \uparrow,\downarrow \downarrow}(t) & 0 &0 &1
\end{pmatrix} \label{rhos}
\end{equation}
where the decoherence factor $D_{\uparrow \uparrow,\downarrow \downarrow}(t)=\langle{\varphi_{\downarrow \downarrow}(t)}|{\varphi_{\uparrow \uparrow}(t)}\rangle$ is explicitly given by  
\begin{equation}
D_{\uparrow \uparrow,\downarrow \downarrow}(t)=\bra{G(\lambda_i)}e^{iH_{\downarrow \downarrow}(\lambda_f)t}e^{-iH_{\uparrow \uparrow}(\lambda_f)t}\ket{G(\lambda_i)} \; .
\end{equation}
Since the populations of the two defect spins do not change in time we see here that our model describes in the computational base a purely dephasing dynamics.

The decoherence factor $D$, governing the dynamics of the qubits, is simply related to the so called Loschmidt echo \cite{gou12} 
via
\begin{equation}
\mathcal{L}_{\uparrow \uparrow,\downarrow \downarrow}(t)=\big|\bra{G(\lambda_i)}e^{iH_{\downarrow \downarrow}(\lambda_f)t}e^{-iH_{\uparrow \uparrow}(\lambda_f)t}\ket{G(\lambda_i)}\big|^2 \; .
\label{echo}
\end{equation}
Notice that if the final magnetic field is equal to the initial one ($\lambda_i=\lambda_f$, meaning that the bath is not quenched), the initial state $\ket{G(\lambda_i)}$ is the ground state of the Hamiltonian $H_{\downarrow \downarrow}(\lambda_i)$ and the echo is reduced to $\mathcal{L}(t)=\big|\bra{G(\lambda_i)}e^{-iH_{\uparrow \uparrow}(\lambda_i)t}\ket{G(\lambda_i)}\big|^2$, which is the case treated in \cite{Cor08}.

\subsection{Entanglement measure}
We use the Wooter's concurrence \cite{woo98,woo01} as the entanglement measure of our qubits system since in such a case it takes a very simple form. For a two-qubits system the concurrence associated with a state $\rho$ is given by
\begin{equation}
\mathcal{C}(\rho)=\textrm{max}\{0, \varepsilon_1-\varepsilon_2-\varepsilon_3-\varepsilon_4\}
\end{equation}
where the $\varepsilon_i$'s are the square roots of the eigenvalues in decreasing order of the (generally) non Hermitian matrix $R=\rho\tilde{\rho}$ with $\tilde{\rho}$ defined as
\begin{equation}
\tilde{\rho}=(\sigma_y \otimes \sigma_y)\rho^*(\sigma_y \otimes \sigma_y)\; ,
\end{equation}
where the complex conjugation is taken in the computational base. 
For the density matrix \eqref{rhos} the matrix $\tilde{\rho}=\rho$ and then  
\begin{equation}
R=\rho^2=\frac{1}{4}
\begin{pmatrix}
1+|D|^2 & 0 & 0 & 2D \\
0 & 0 & 0 &0 \\
0 & 0 & 0 &0 \\
2D^* & 0 & 0 & 1+|D|^2 \\
\end{pmatrix} \; ,
\end{equation}
which leads for the eigenvalues to $\varepsilon_1=\frac{1}{4}(1+|D|)^2, \varepsilon_2=\frac{1}{4}(1-|D|)^2, \varepsilon_3=\varepsilon_4=0$. Finally, for the state \eqref{rhos} the concurrence is simply given by
\begin{equation}
\mathcal{C}_{AB}(t)=\sqrt{\mathcal{L}(t)}=|D(t)|\; .
\end{equation}
The entire dynamics of the two qubits $A$ and $B$ is encoded in the Loschmidt echo and the main goal of this study is then to determine it.

\section{Loschmidt echo in the Fermionic representation}
\subsection{Jordan Wigner transformation}
The dynamics of the qubits system is generated through the two environment channels described by $H_{\uparrow \uparrow}$ and $H_{\downarrow \downarrow}$ which, as stated before, have the same structure except for two defects transverse fields at positions $0$ and $d$. Apart from that, these Hamiltonians are both diagonalized through the same standard procedure, that is performing a Jordan-Wigner mapping followed by a Bogoliubov transformation and in the following we drop out the indices ${\uparrow \uparrow},{\downarrow \downarrow}$. 
In terms of the ladder operators $\sigma^{\pm}=\frac{\sigma^x \pm i\sigma^y}{2}$ the Jordan-Wigner mapping reads
\be
\sigma_j^+=e^{i\pi \sum_{i=0}^{j-1}c_i^\dagger c_i}c_j^\dagger \; , \quad \sigma_j^-=c_je^{-i\pi \sum_{i=0}^{j-1}c_i^\dagger c_i}\; ,
\ee
where the operators $c_j$ and $c_j^\dagger$ satisfy the canonical Fermi algebra $\{c_i,c_j^\dagger \}=\delta_{i,j}$\; ,  
$\{c_i,c_j \}= \{c_i^\dagger,c_j^\dagger \}=0$. 
In terms of the Fermi algebra the environment Hamiltonians in the relevant parity sector become
\begin{equation}
H=\sum_{i,j}(c_i^\dagger A_{ij}c_j+\frac{1}{2}(c_i^\dagger B_{ij}c_j^\dagger+h.c.))\; ,
\end{equation}
with $A_{ij}=-2\lambda_i \delta_{ij} -J[\delta_{i,j-1}+\delta_{i,j-1}]$ and $B_{ij}=J[ \delta_{i,j+1}-\delta_{i,j-1}]$ (indices $N$ are identified with $0$ to account for the periodic boundaries) defining respectively $N\times N$ symmetric and antisymmetric matrices $A$ and $B$. 
Introducing the field operator 
\be
\boldsymbol{\Psi}^\dagger=(\boldsymbol{C},\boldsymbol{C}^\dagger)=(c_0, \hdots, c_{N-1},c_0^\dagger, \hdots, c_{N-1}^\dagger)\; ,
\ee
the Hamiltonian is further rewritten in a more compact form
\begin{equation}
H=\frac{1}{2}\boldsymbol{\Psi}^\dagger {\mathcal{H}}\boldsymbol{\Psi},
\end{equation}
with the single particle Hamiltonian
\begin{equation}
{\mathcal{H}}=
\begin{pmatrix}
-A & -B \\
B & A
\end{pmatrix} \; .
\label{H}
\end{equation}
In order to diagonalize the Hamiltonian $H$ we introduce the unitary matrix 
\begin{equation}
U=
\begin{pmatrix}
g & h \\
h & g
\end{pmatrix}
\end{equation}
that diagonalizes the single particle matrix $\mathcal{H}$: $\Lambda =U^\dagger \mathcal{H}U$. 
The Hamiltonian $H$ is readily diagonalized in terms of normal modes $\boldsymbol{\eta}=U^\dagger \boldsymbol{\Psi}$ and takes the form
\be
H=\frac{1}{2}\boldsymbol{\eta}^{\dagger}\Lambda\boldsymbol{\eta}\; .
\label{Hfield}
\ee
More explicitly, the normal modes operators $\boldsymbol{\eta}^{\dagger}=(\eta_0, \hdots, \eta_{N-1}, \eta^{\dagger}_0, \hdots, \eta^{\dagger}_{N-1} )$ are related to the original fermi operators by the real Bogoliubov coefficients $g_{ij}$ and $h_{ij}$ through
\be
\eta_k=\sum_{i}(g_{ik}c_i+h_{ik}c_i^{\dagger}) 
\ee
and similar expressions for the adjoins $\eta_k^\dagger$. These relations are easily inverted and lead to
\be
c_i=\sum_{k}(g_{ik}\eta_k+h_{ik}\eta_k^{\dagger}) 
\ee
for the original Fermi operators in terms of the normal modes operators.

\subsection{Time evolution of the covariance matrix and Loschmidt echo}
Since the Hamiltonians $H_\alpha$ with  $\alpha=\uparrow \uparrow, \downarrow \downarrow$ are free fermionic, 
the Loschmidt echo \eqref{echo}, describing the overlap between the states $\ket{\varphi_{\alpha}(t)}=e^{-iH_\alpha t}\ket{G(\lambda_i)}$, can be expressed in terms of the covariance matrices 
\be
C_\alpha(t)= \bra{\varphi_{\alpha}(t)}\boldsymbol{\Psi}\boldsymbol{\Psi^\dagger}\ket{\varphi_{\alpha}(t)}
\ee
only and reads \cite{key10}
\be
\mathcal{L}_{\uparrow \uparrow,\downarrow \downarrow}(t)={\big|\det\big(\mathbb{1}-C_{\downarrow \downarrow}(t)-C_{\uparrow \uparrow}(t)\big)\big|}^{1/2}\; ,
\label{echofinal}
\ee
where $\mathbb{1}$ is the $2N\times 2N$ identity matrix. The problem of computing the Loschmidt echo is then related to the evaluation of the time-evolved covariance matrices $C_\alpha (t)$. 
In order to derive this time dependence it is more convenient to switch to the Heisenberg picture. Thanks to the quadratic structure of the Hamiltonians $H_\alpha$, the equations of motion for the field operators $\boldsymbol{\Psi}_\alpha$ in each channels $\alpha=\uparrow \uparrow, \downarrow \downarrow$ take the form
\be
i\frac{d}{dt}\boldsymbol{\Psi}_\alpha=\mathcal{H}_\alpha \boldsymbol{\Psi}_\alpha\; ,
\ee
where $\mathcal{H}_\alpha$ is the single particle Hamiltonian \eqref{H} associated to the channel $\alpha$. 
Together with the initial conditions $\boldsymbol{\Psi}_\alpha(0)=\boldsymbol{\Psi}$, these equations of motion are easily integrated and lead to 
$\boldsymbol{\Psi}_\alpha(t)=e^{-it\mathcal{H}_\alpha}\boldsymbol{\Psi}$. 
This allows us to write the time evolution of the covariance matrix as
\begin{equation}
C_\alpha(t)=e^{-it\mathcal{H}_\alpha}C(0)e^{it\mathcal{H}_\alpha}\; ,
\label{Ct}
\end{equation}
with $C(0)=\bra{G(\lambda_i)}\boldsymbol{\Psi}\boldsymbol{\Psi^\dagger}\ket{G(\lambda_i)} $  the initial covariance matrix. In terms  of the field operators $\boldsymbol{C}$ and $\boldsymbol{C}^\dagger$ it is given by
\begin{align}
C(0)=\begin{pmatrix}
\langle \boldsymbol{C}^\dagger  \boldsymbol{C} \rangle & \langle  \boldsymbol{C}^\dagger  \boldsymbol{C}^\dagger \rangle \\
\langle  \boldsymbol{C} \boldsymbol{C} \rangle & \langle  \boldsymbol{C} \boldsymbol{C}^\dagger \rangle 
\end{pmatrix}\; ,
\end{align}
where $\langle . \rangle$ is the operators expectation value in the ground state $\ket{G(\lambda_i)}$. 
Consequently, the Loschmidt echo \eqref{echofinal} is explicitly derived from \eqref{Ct} given the initial covariance matrix $C(0)$.

\section{Quench dynamics}

\subsection{Weak and strong coupling regimes}
\begin{figure}[h!]
\begin{center}
\includegraphics[scale=0.3]{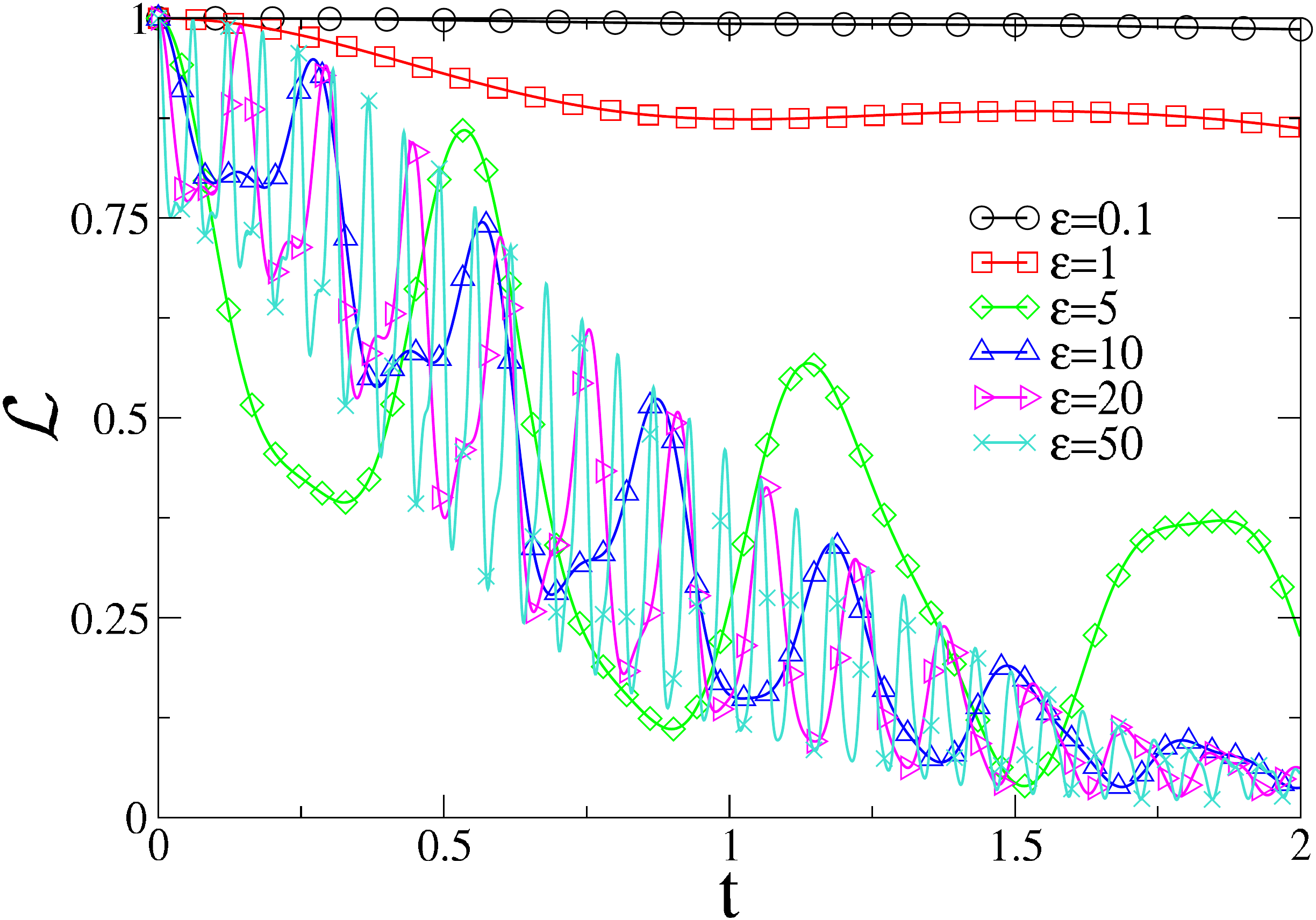}
\caption{(Color online) Time evolution of the Loschmidt echo after a quench from $\lambda_i=1.5$ to $\lambda_f=0.5$ for several coupling strengths $\varepsilon$. The distance is fixed to $d=1$ and the size of the environment is $N=100$.}
\label{fig_epsilon}
\end{center}
\end{figure}

Let us consider first the influence of the coupling strength $\varepsilon$ on the decoherence dynamics of the qubits for a given quench protocol. In figure \ref{fig_epsilon}  we have plotted the time evolution of the Loschmidt echo as a function of $\varepsilon$ for an initial field $\lambda_i=1.5$ and quenched at $\lambda_f=0.5$.
One sees that at a given quench protocol the decoherence is faster when the coupling strength is increased. Whereas the echo decreases slowly for weak coupling $\varepsilon\ll1$, the behavior is quite different in the strong coupling regime $\varepsilon\gg1$. Indeed, one observes fast oscillations of the echo $\mathcal{L}$ which are embedded inside an envelope which is independent of the coupling strength at sufficiently large $\varepsilon$ ($\varepsilon\ge 10$ in figure  \ref{fig_epsilon}). Note that this effect is not a consequence of the quench in the chain, since it has already been observed in the equilibrium situation $\lambda_i=\lambda_f$ as well. These fast oscillations are directly related to the two high frequencies, proportional to the coupling strength $\varepsilon$, generated by the coupling of the qubits to the chain, whereas the remaining smaller frequencies (independent of $\varepsilon$) are responsible of the slower decay of the envelope.

\subsection{Effect of the quench on the Loschmidt echo}
\begin{figure}[h!]
\begin{center}
\includegraphics[scale=0.33]{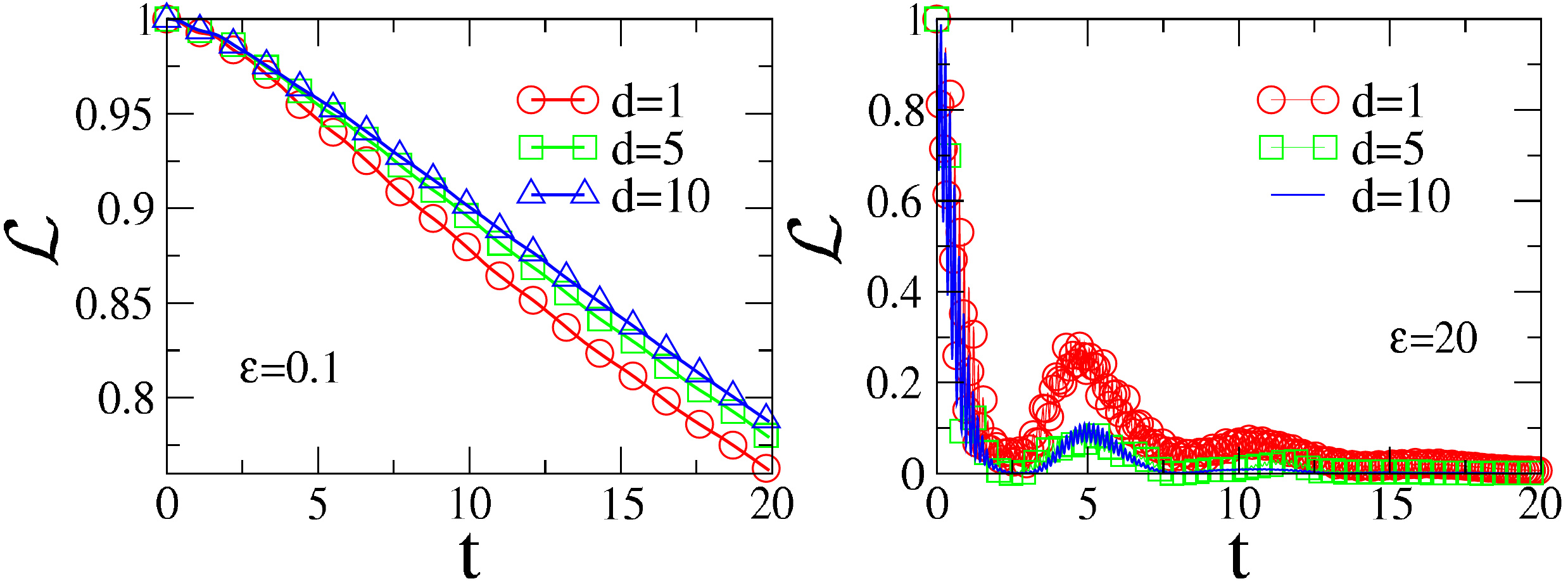}
\caption{(Color online) Time evolution of the Loschmidt echo for distances $d=1$, $d=5$, and $d=10$ in the weak (left) and strong (right) coupling regimes. The quench is done from $\lambda_i=1.5$ to $\lambda_f=0.5$ and the chain is made of $N=100$ spins.}
\label{fig_distance}
\end{center}
\end{figure}

We first analyze roughly the effect of the sudden quench dynamics through the evolution of the Loschmidt echo  obtained from \eqref{echofinal} and \eqref{Ct} by exact numerical diagonalization.
Figures~\ref{echo} and  \ref{echo_sc} show the time evolution of the Loschmidt echo for several quench protocols for an Ising chain of fixed size $N=100$, $J=1$, distance $d=1$ between the qubits and coupling constants $\varepsilon=0.1$ and $\varepsilon=20$ respectively.
The first observation that can be made is that the decoherence (and then the disentanglement) is enhanced at large times by the quench in comparison to  the un-quenched situation (full lines in figure \ref{echo} and red curves in figure \ref{echo_sc}), 
for both weak and strong coupling regime. We also notice that bigger the quench amplitude $|\lambda_f-\lambda_i|$  becomes, stronger the disentanglement becomes. This phenomenon is observed numerically whatever the distance between the qubits is. The behavior of the echo with the qubits distance is opposite in weak and strong coupling regimes: for weak coupling, the echos decreases with the distance whereas it increases with the distance in the strong coupling regime \cite{Cor08}, as it can be seen on figure  \ref{fig_distance} where we show the time evolution of the echo for different distances in the two coupling regimes.
Moreover, one observes in the weak coupling regime that the decrease of the Loschmidt echo is monotonous during the time evolution apart for small superimposed oscillations. One can observe beating of the envelope in the strong coupling regime, see for example the red curves of plots a) and c) figure \ref{echo_sc}. This phenomenon, already observed at equilibrium in \cite{Cor08}, can be explained in terms of a decomposition of the spectrum of the Hamiltonian. Indeed, as we mentioned previously, the strong coupling of the qubits to the chain brings two high frequency excitations of the order of $\varepsilon$, whereas the remaining part of the spectrum can be split into two regions corresponding respectively to the region lying between the two qubits and the region lying outside the interaction sites (this decomposition make sense since $d\ll N$). For fields smaller than the critical field, it turns out that 
the beating observed in the echo is associated to the lowest energy excitations of the region between the qubits \cite{Cor08}. 
When the magnetic field increases above the critical value, more and more modes start to be populated leading to the disappearance of the phenomenon.

In order to characterize the effect of the sudden quench on the disentanglement, we will use this monotonic decrease of $\mathcal{L}$ in the weak coupling regime.
 We have plotted in the left panel of figure~3 the Loschmidt echo at a fixed large enough time ($t=10$) as a function of the initial transverse field value $\lambda_i$ at two fixed post-quench values $\lambda_f$ (bellow and above the critical value $\lambda_c=1$) and in the right panel the echo at the same time as a function of the final field at fixed initial fields. 
We see clearly on these figures that the echo presents a maximum value at the un-quenched point (equilibrium situation $\lambda_i=\lambda_f$)  showing that the non-equilibrium situation ($\lambda_i\neq \lambda_f$) is always unfavorable with respect to the coherence of the qubits.
At the equilibrium point, one recovers the value of $\mathcal{L}$ already found in \cite{Cor08}. 
Away from it, one observes that in the large field limit the Loschmidt echo saturates at a constant value. This saturation of the decoherence for high initial magnetic field is easy to understand. Indeed, if $\lambda_i$ is very large, the initial state is close to a completely polarized state along the direction of the field $\ket{\Psi}=\ket{\uparrow \uparrow \hdots \uparrow}$. In this limiting case, the initial covariance matrix is trivially
\begin{equation}
C(0)=
\begin{pmatrix}
\mathbb{1} & 0\\
0 & 0
\end{pmatrix}
\end{equation} \\
and obviously does not depend anymore on the initial magnetic field $\lambda_i$ and consequently neither does the Loschmidt echo. 
In the left panel of figure~3 the saturation value of the echo for a completely polarized initial state is shown in dashed lines for the two different final fields considered there. We see that the Loschmidt echo converges asymptotically to these limiting values.  
On the right panel of figure~\ref{lambda}, one sees that the same saturation phenomenon applies with respect to large final fields. 

\begin{figure}[h!]
\begin{center}
\includegraphics[scale=0.33]{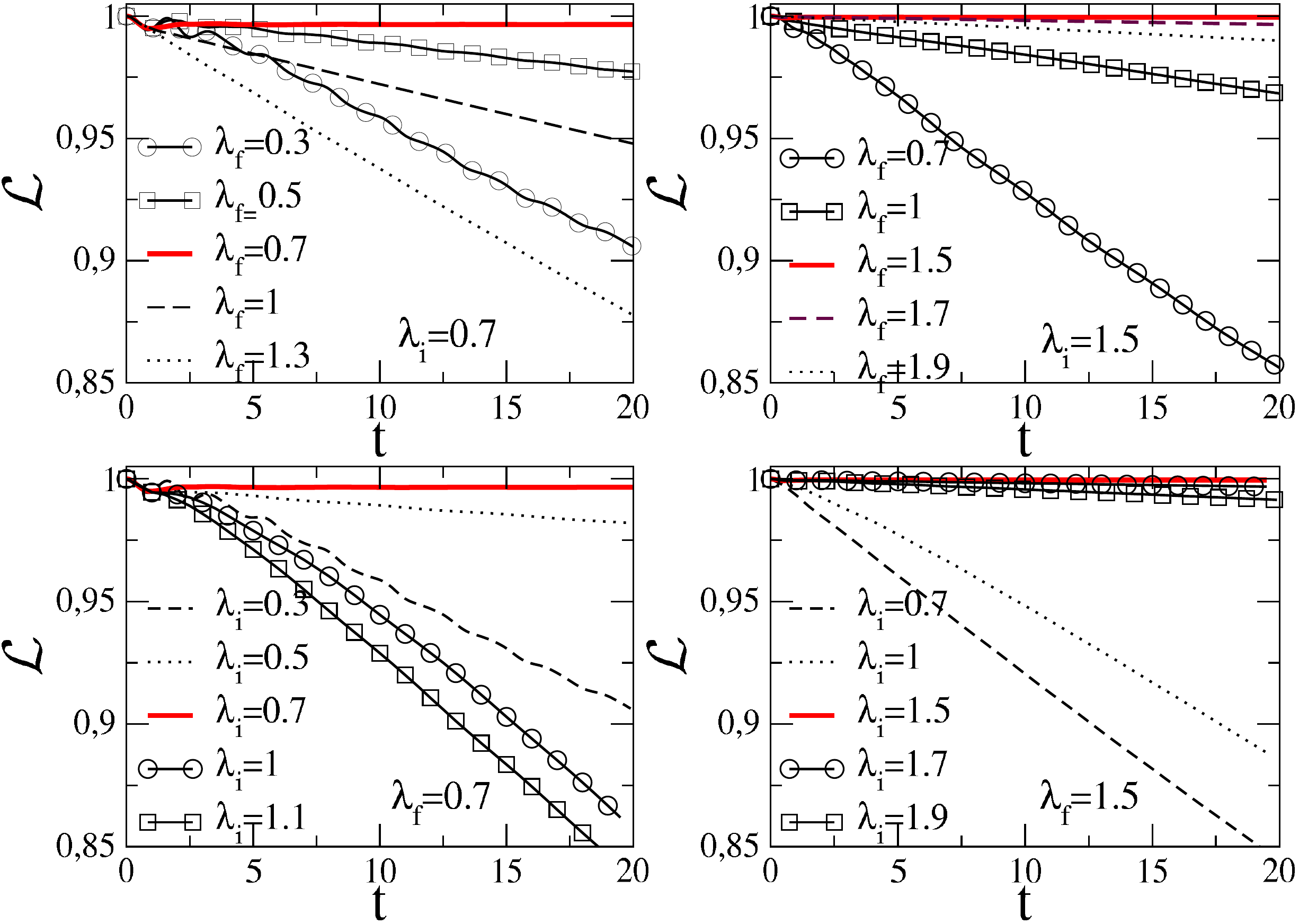}
\caption{(Color online) Time evolution of the echo in the weak coupling regime for different quench protocols. For all plots, we choose $N=100$, $\varepsilon=0.1$ and keep fixed the distance to $d=1$. The two up plots are a variation of the final magnetic field whereas the two down plots are a variation of the initial one. For all plots, the varied field is plotted with symbols for $\lambda_i>\lambda_f$, with dashed line for $\lambda_i<\lambda_f$ and in full line in the equilibrium case $\lambda_i=\lambda_f$.}
\label{echo}
\end{center}
\end{figure}

\begin{figure}[h!]
\begin{center}
\includegraphics[scale=0.33]{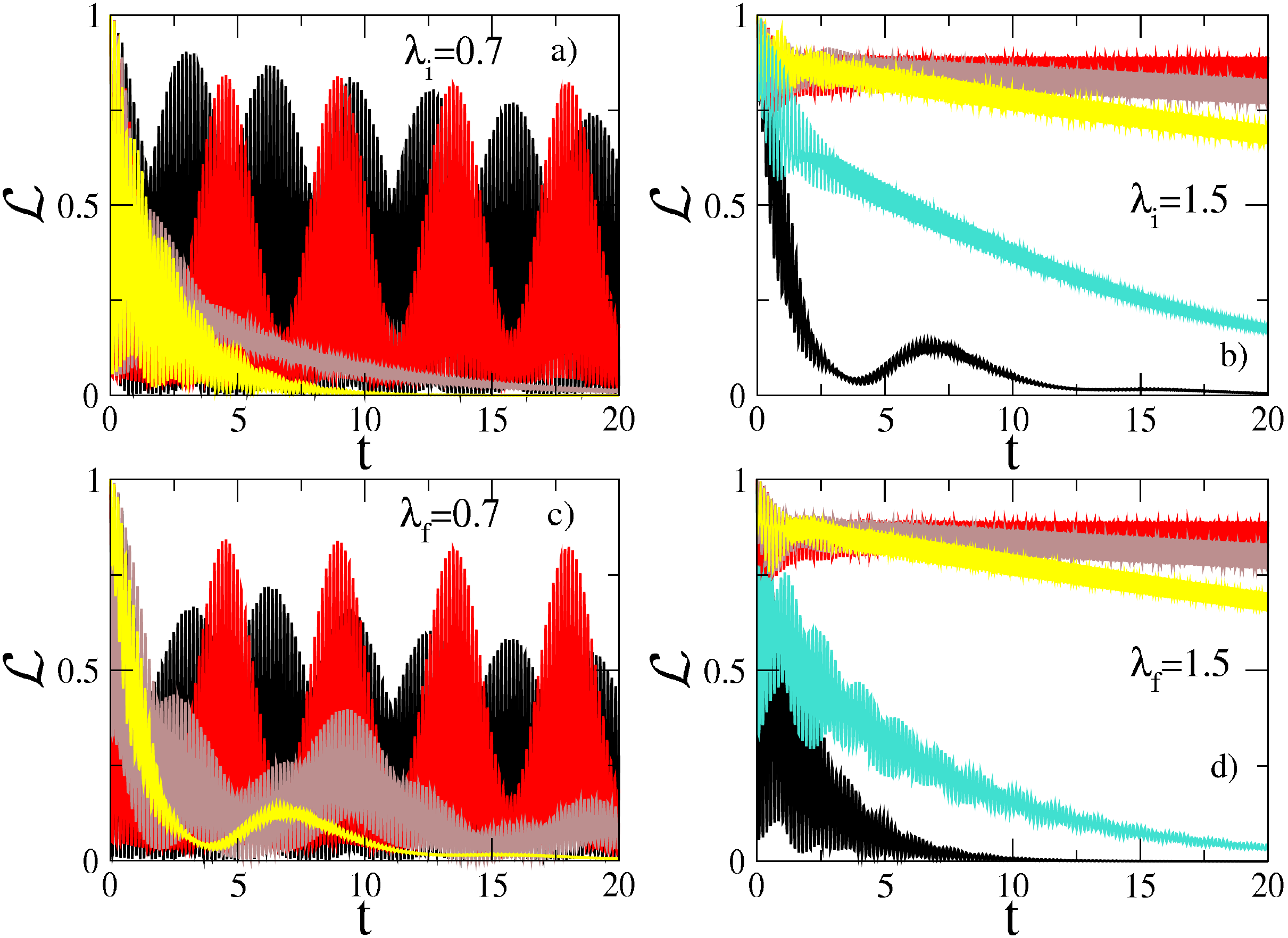}
\caption{(Color online) Time evolution of the echo in the strong coupling regime for different quench protocols. For all plots, we choose $N=100$, $\varepsilon=20$ and keep fixed the distance to $d=1$. The two plots a) and c) are a variation of the final magnetic field whereas plots b) and d) are a variation of the initial one. For a) and c), the varied fields are $\lambda_{i,f}=0.5$ (black), $\lambda_{i,f}=0.7$ (red), $\lambda_{i,f}=1$ (brown) and $\lambda_{i,f}=1.5$ (yellow). For b) and d), the varied fields are, $\lambda_{i,f}=0.7$ (black), $\lambda_{i,f}=1$ (light blue), $\lambda_{i,f}=1.5$ (red), $\lambda_{i,f}=1.7$ (brown) and $\lambda_{i,f}=1.9$ (yellow).}
\label{echo_sc}
\end{center}
\end{figure}

\begin{figure}[h!]
\begin{center}
\includegraphics[scale=0.35]{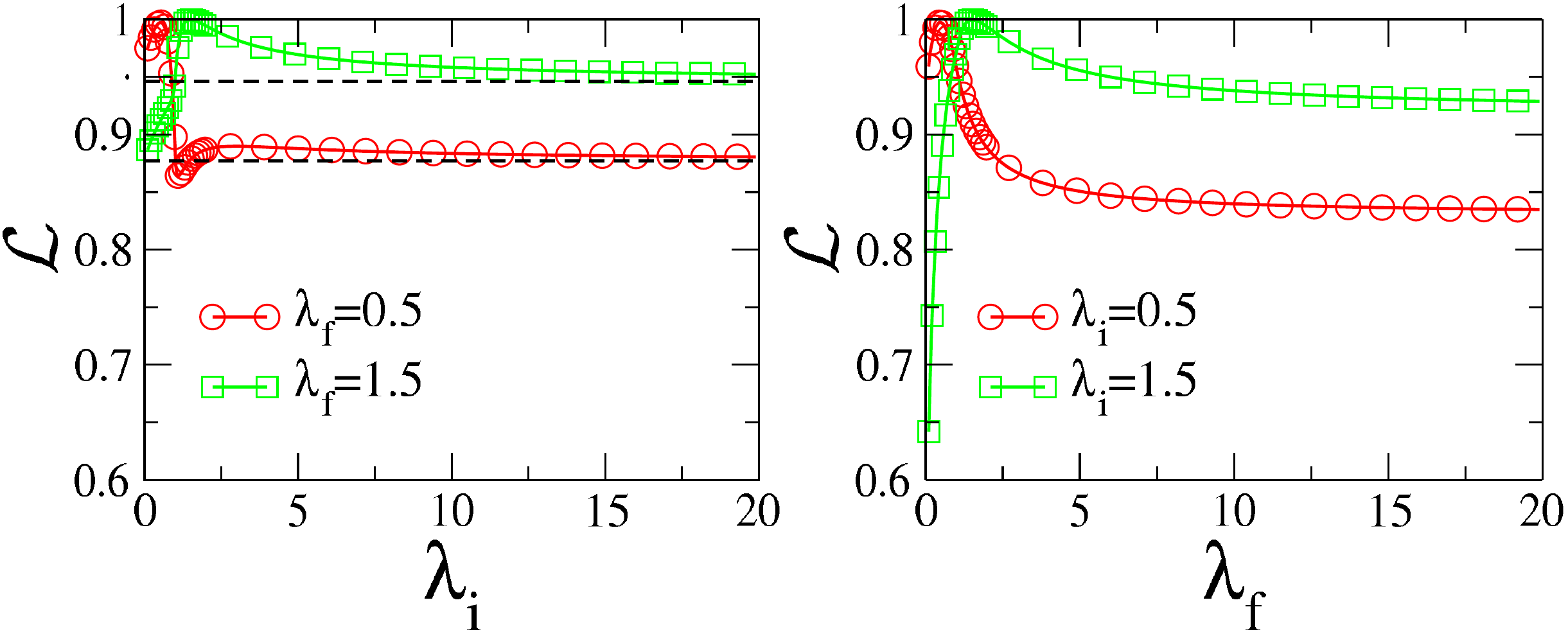}
\caption{(Color online) Loschmidt echo at time $t=10$ as a function of the initial (final fixed) (left) and final (initial fixed) (right) magnetic field. The varied magnetic fields are $0.5$ (red circles) and $1.5$ (green squares). The dashed lines represent the limiting case of a completely polarized initial state ($J=0$).}
\label{lambda}
\end{center}
\end{figure}

\noindent

\begin{figure}[h!]
\begin{center}
\includegraphics[scale=0.4]{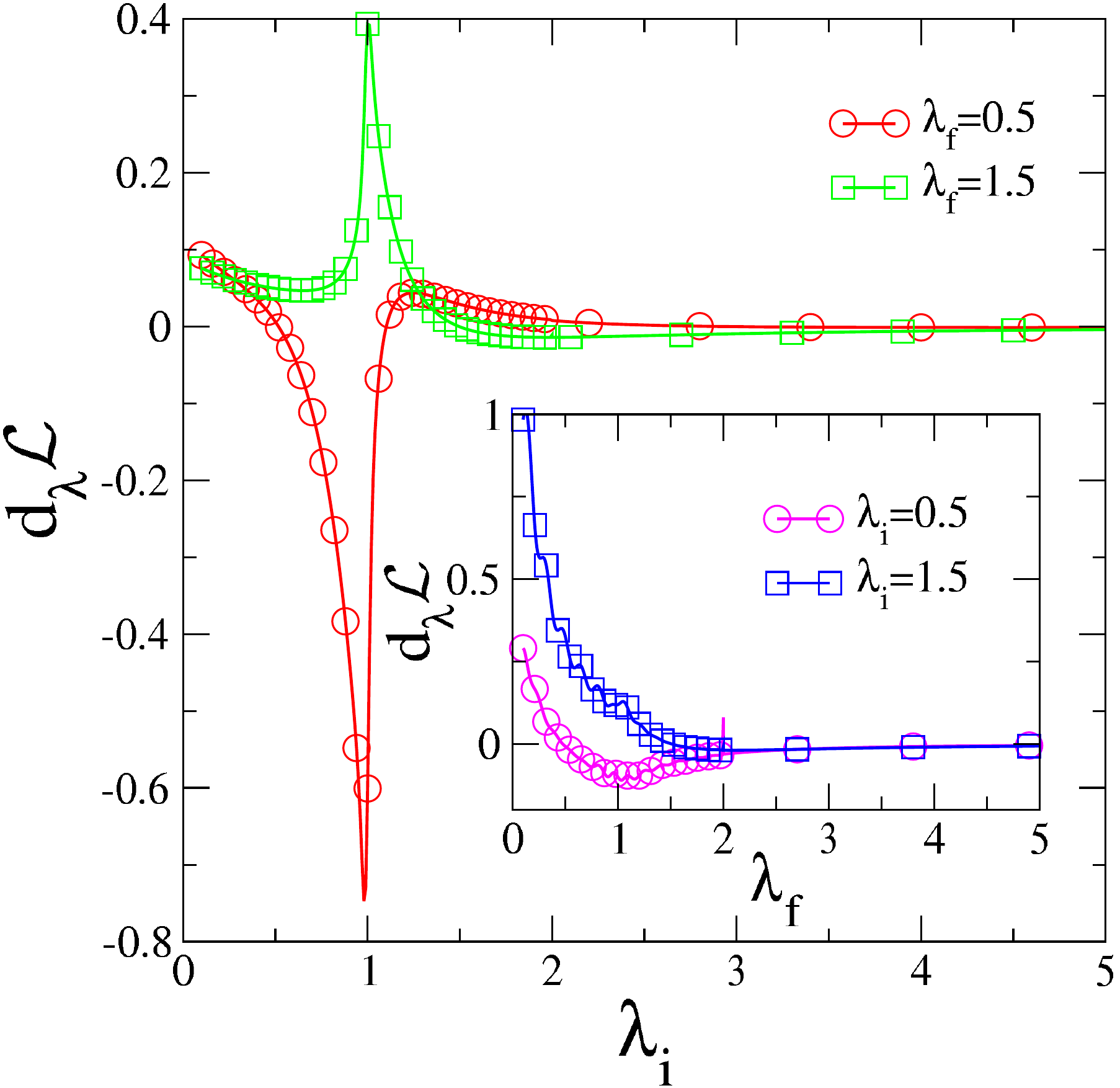}
\caption{(Color online) First derivative of the Loschmidt echo at time $t=10$ as a function of the initial magnetic field for $\lambda_f=0.5$ (red circles) and $\lambda_f=1.5$ (green squares). Other parameters are $\varepsilon=0.1$, $d=1$ and $N=100$. In the inset, we show the derivative of $\mathcal{L}$ with respect to the final field $\lambda_f$ for $\lambda_i=0.5$ (magenta circles) and $\lambda_i=1.5$ (blue squares).}
\label{der_lambda}
\end{center}
\end{figure}

\noindent
The Loschmidt Echo, and subsequently the entanglement, exhibits a signature of the quantum phase transition experienced by the Ising chain. Indeed, when the initial magnetic field is varied, we clearly see a jump in the curve for $\lambda_i$ close to the critical value $\lambda_c=1$. This critical behavior is better seen by analyzing the first derivative of $\mathcal{L}(t)$ with respect to the initial field $\lambda_i$. The derivative with respect to the initial field,  at fixed time $t=10$, is plotted in figure \ref{der_lambda} for final fields in the ordered ($\lambda_f=0.5$) and disordered phase ($\lambda_f=1.5$). 
For the two cases, the first derivative exhibits a clear singularity when the bath approaches criticality. Notice that on one hand the derivative is negative  for $\lambda_f=0.5$ reflecting the fact that the divergence occurs after the equilibrium point ($\lambda_f=\lambda_i=0.5$) when the Echo is decreasing with the field. On the other hand, it is positive at $\lambda_f=1.5$, since the divergence occurs before the equilibrium point ($\lambda_f=\lambda_i=1.5$) when the echo is increasing with the field.
On the other side, there is no clear signature of a singularity, as seen from the inset of figure \ref{der_lambda}, with respect to a variation of the final field for fixed initial fields $\lambda_i=0.5$ and $\lambda_i=1.5$. This indicates that the critical behavior is totally set by the initial state of the environment, whereas the final magnetic field is only responsible for dynamical effects, as we will see latter.

\begin{figure}[h!]
\begin{center}
\includegraphics[scale=0.3]{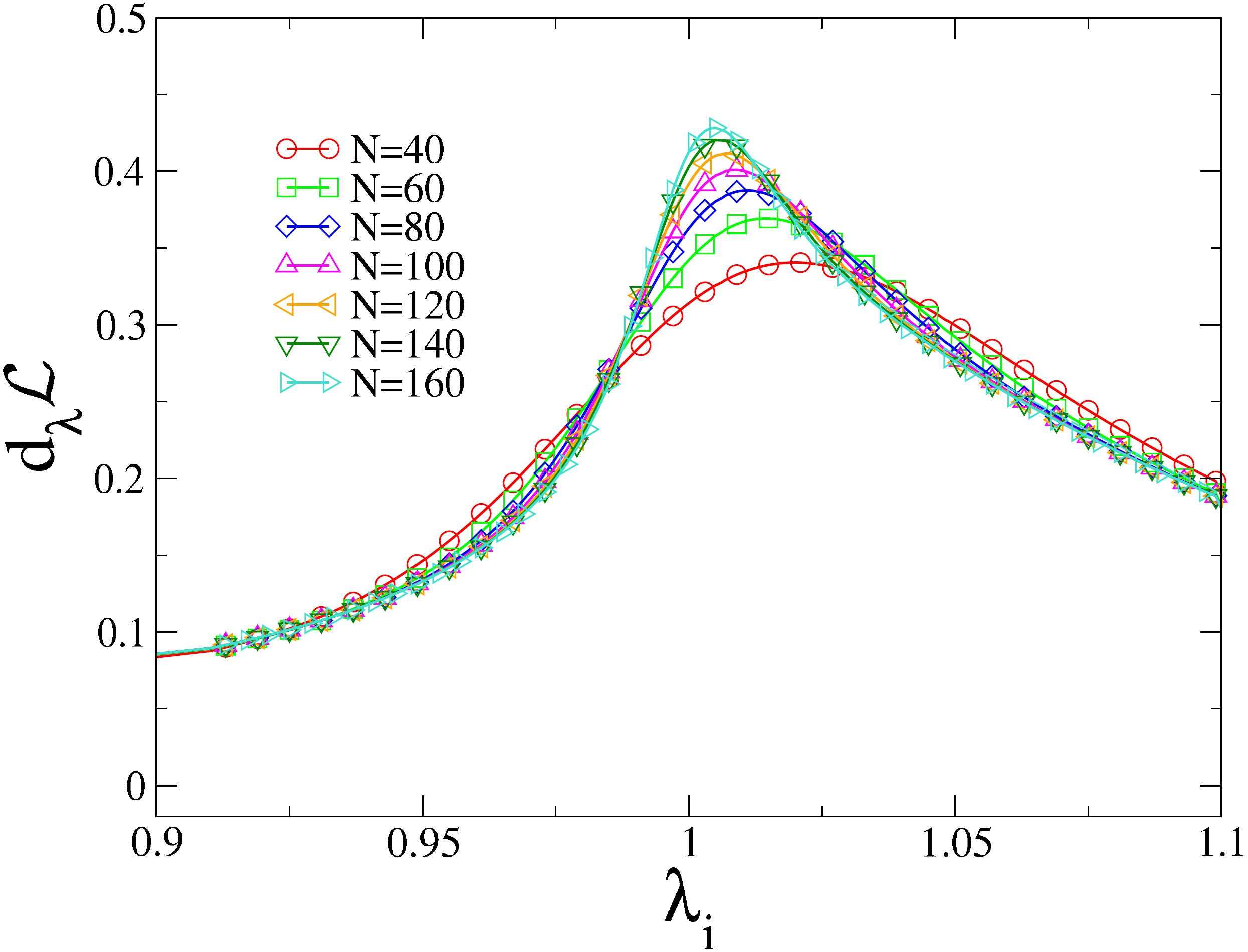}
\caption{(Color online) First derivative of the Loschmidt echo at time $t=10$ as a function of the initial magnetic field for different sizes of the bath with $\lambda_f=1.5$, $\varepsilon=0.1$ and $d=1$.}
\label{der_lambda_size}
\end{center}
\end{figure}

\begin{figure}[th]
\begin{center}
\includegraphics[scale=0.33]{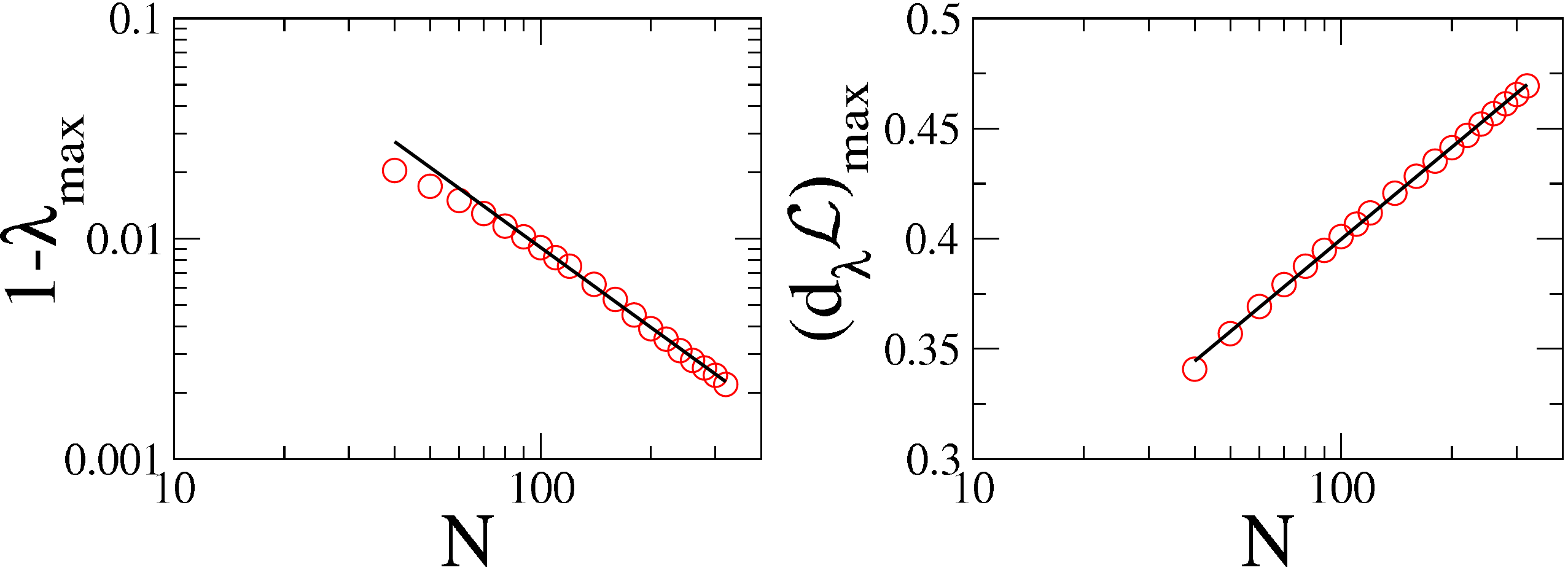}
\caption{(Color online) Left: Scaling behavior of the position of the peaks $\lambda_{max}$ as a function of the size of the bath $N$. 
Right:  Scaling behavior of the maximum value reaches by $d_{\lambda}\mathcal{L}$ as a function of the size of the bath $N$. 
Parameters are $\lambda_f=1.5$, $\varepsilon=0.1$ and $d=1$.}
\label{scaling}
\end{center}
\end{figure}

Due to the finite size of the environment, the singularity of the absolute value of the first derivative of the Loschmidt echo is rounded and reaches a maximum value at a given $\lambda_{max}$ of the initial  field, see figure  \ref{der_lambda_size}. 
As the size of the environment increases, the maximum value of $|d_{\lambda}\mathcal{L}|$ diverges logarithmically with the size: $ d_{\lambda}\mathcal{L}|_{\lambda_{max}} \sim \ln N$. At the same time the value $\lambda_{max}$ of the initial field  approaches asymptotically the critical value $\lambda_c=1$ as $|\lambda_c-\lambda_{max}|\sim N^{\gamma}$ with an exponent $\gamma$ which is found numerically to be $\sim-1.1$, as shown in figure \ref{scaling}. The expected value from critical scaling theory \cite{Hen99} is $\gamma=-1/\nu=-1$, since the correlation length exponent $\nu=1$ for the quantum Ising chain. The departure from that value is due to quite strong corrections to scaling and is numerically compatible with a scaling correction 
$N(\lambda_c-\lambda_{max})\sim 1 +const./N$.  
Notice that these scaling results are coherent with those found in Ref\cite{ost02,zho08}.

\subsection{Short time behavior}
\begin{figure}[h!]
\begin{center}
\includegraphics[scale=0.34]{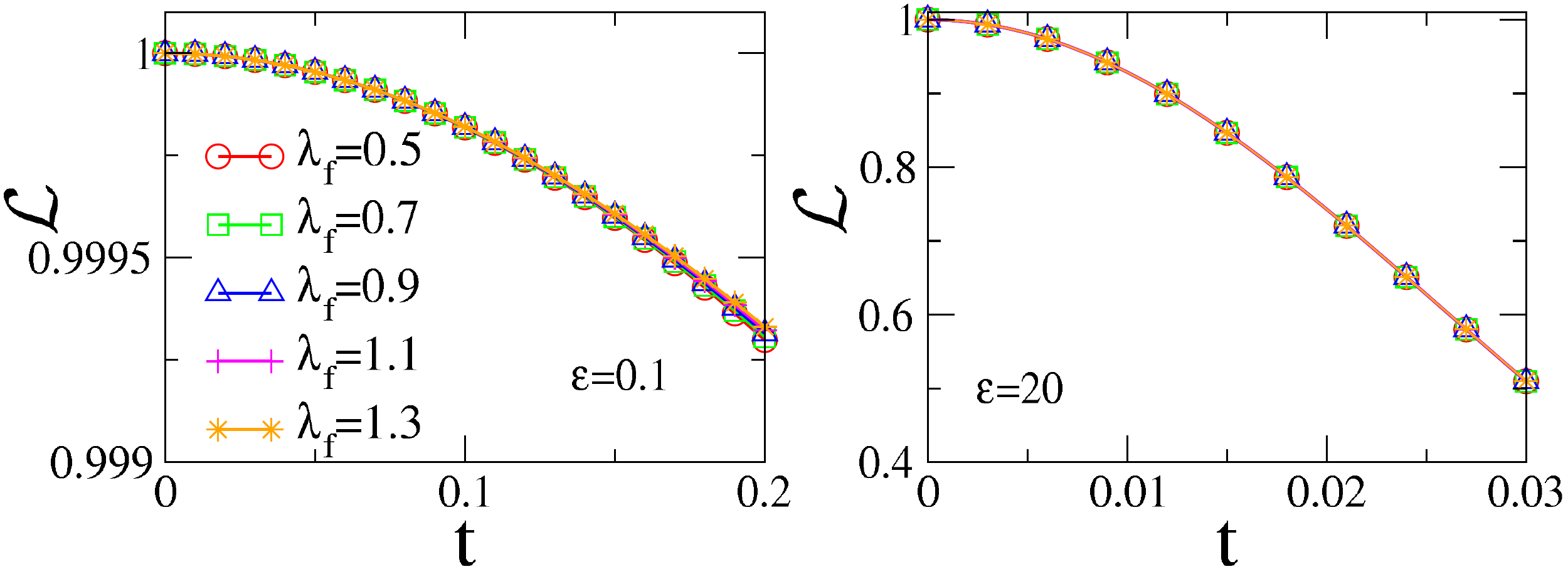}
\caption{(Color online) Short time evolution of the Loschmidt echo for different values of the final field and $\varepsilon=0.1$ (left) and $\varepsilon=20$ (right). Other parameters are $\lambda_i=0.7$ and $N=100$.}
\label{short_time}
\end{center}
\end{figure}

For times much shorter than the typical time scale of the system $t\ll t_{typ}$ with
\begin{align}
t_{typ}=&1 \quad \textrm{for}\ \varepsilon\ll1,\\
t_{typ}=&1/\varepsilon \quad \textrm{for}\ \varepsilon\gg1,
\end{align}
the Loschmidt echo shows a parabolic decay independent of the quench parameters as seen on figure 
\ref{short_time}. 
This independence is easily understood from a perturbative approach  \cite{per84}. Indeed, 
expanding the ground state $\ket{G(\lambda_i)}$ in the eigenbasis $\{\ket{\phi_m}\}$ and $\{\ket{\varphi_m}\}$ of 
$H_{\downarrow \downarrow}$ and $H_{\uparrow \uparrow}$ respectively, $\ket{G(\lambda_i)}=\sum_m a_m \ket{\phi_m} =
\sum_m b_m \ket{\varphi_m}$,
the echo becomes
\be
\mathcal{L}(t)=\left|\sum_{mn}a_m^*b_ne^{-i(E^{\uparrow \uparrow}_n-E^{\downarrow \downarrow}_m)t}\langle\phi_m|\varphi_n\rangle\right|^2\; .
\ee
At first order in perturbation theory, the eigenvalues are given by 
\begin{equation}
E^{\uparrow \uparrow}_n=E^{\downarrow \downarrow}_n+\bra{\phi_n}\tilde{H}_I\ket{\phi_n}=E^{\downarrow \downarrow}_n+V_n \; .
\end{equation}
where $\tilde{H}_I=-\varepsilon(\sigma_0^z+\sigma_d^z)$. If the interaction Hamiltonian is sufficiently small,  the decomposition coefficients  $a_m\approx b_m$ and  $\langle\phi_m|\varphi_n\rangle\approx \delta_{m,n}$ such that
\begin{equation}
\mathcal{L}(t)\approx \left|\sum_{n}|a_n|^2e^{-iV_n t}\right|^2 \; .
\end{equation}
Expanding the exponential up to second order in time one obtains 
\begin{align}
\mathcal{L}(t) \approx& \left|\sum_{m}|a_m|^2\left(1-itV_m-\frac{t^2}{2}(V_m)^2\right)\right|^2 \nonumber \\
\approx& 1-\left( \sum_m |a_m|^2 V_m^2 -\left(\sum_m |a_m|^2 V_m\right)^2\right)t^2 \nonumber \\
\approx& 1-\left(\langle\tilde{H}_I^2\rangle-\langle\tilde{H}_I\rangle^2 \right)t^2 \equiv 1-\alpha t^2 \; .
\label{alpha}
\end{align}
Then, for short times, the echo depends only on the variance of the interaction Hamiltonian over the initial state $\ket{G(\lambda_i)}$ and consequently not on the quench protocol itself.

\begin{figure}[h!]
\begin{center}
\includegraphics[scale=0.335]{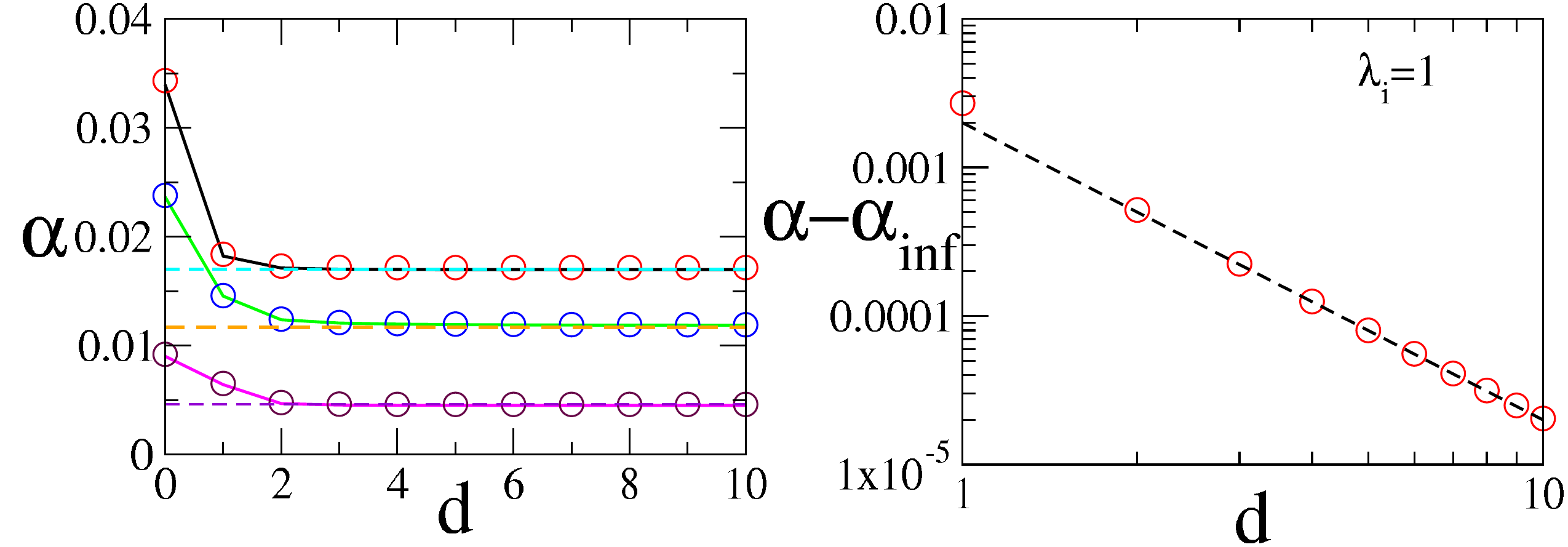}
\caption{(Color online) Gaussian rate $\alpha$ as a function of the parameters of the system. Left: as a function of the distance $d$ for a final field $\lambda_f=0.5$ and, from top to bottom, $\lambda_i=0.7$, $\lambda_i=1$ and $\lambda_i=1.5$. The circles are the numerical fits of the echo, the full lines are obtained with equation \eqref{alpha_theo}, and the horizontal dashed lines give the asymptotic ($d\rightarrow \infty$) values of $\alpha$. 
Right: $\alpha-\alpha_{d\rightarrow \infty}$ in the critical case $\lambda_i=1$ and for $\lambda_f=0.5$ as a function of the distance $d$ showing a power-law behavior with exponent $-2$ shown in dashed curve. }
\label{parameter}
\end{center}
\end{figure}

The Gaussian rate (the variance) $\alpha$ is easily evaluated by expressing $\tilde{H}_I$ in terms of the normal modes of the Hamiltonian $H_E(\lambda_i)$:
\begin{align}
\tilde{H}_I=&-2\varepsilon \sum_{kl} \left[ (g_{0k}\eta_k^{\dagger}+h_{0k}\eta_k)(g_{0l}\eta_l+h_{0l}\eta_l^{\dagger})\right. 
 \nonumber \\
&  + \left. (g_{dk}\eta_k^{\dagger}+h_{dk}\eta_k)(g_{dl}\eta_l+h_{dl}\eta_l^{\dagger})\right] + 2 \varepsilon.
\end{align}
Using the fact that $\langle\eta_k\eta_l\rangle=\langle\eta_k^{\dagger}\eta_l^{\dagger}\rangle=0$ and $\langle\eta_k\eta_l^{\dagger}\rangle=\delta_{kl}$, the variance $\alpha$ is expressed as
\begin{align}
\alpha=&4\varepsilon^2\sum_{k\ne l}\bigg[(g_{0k}h_{0l} + g_{dk}h_{dl})^2  - 2h_{dk}h_{0l}g_{dl}g_{0k}\nonumber \\
& - h_{0k}h_{0l}g_{0k}g_{0k}- h_{dk}h_{dl}g_{dk}g_{dk}\bigg]\; .
\label{alpha_theo}
\end{align}
Notice that $\alpha$ is nothing but $2 \varepsilon^2 (\langle \sigma_0^z\sigma_{d}^z\rangle_c +1-\langle \sigma^z_0\rangle^2)$ where $\langle AB\rangle_c\equiv \langle AB\rangle-  \langle A\rangle \langle B \rangle$ is the connected correlation function \footnote{Notice also that if we set $d=0$ into this expression (the two spins are coupled at the same site) we recover the formula obtained in \cite{ros07}, but with a coupling constant two times stronger since $\tilde{H}_I=-\varepsilon(\sigma_0^z+\sigma_{d=0}^z)=-(2\varepsilon )\sigma_0^z$.}. In particular, at large distances compared to the correlation length $\xi$ in the initial ground state, i.e. $d\gg \xi$, since $\langle \sigma_0^z\sigma_{d}^z\rangle_c=0$ one expects a saturation value $\alpha(d\gg 1)=2 \varepsilon^2 (1-\langle \sigma^z_0\rangle^2)$. However, when the initial state is critical, that is for $\lambda_i=1$, since the decay of the connected part is algebraic with $\langle \sigma_0^z\sigma_{d}^z\rangle_c\sim d^{-2}$ \cite{Hen99}, the approach toward the saturation value $\alpha(d\rightarrow \infty)$ is algebraic, as shown on figure  \ref{parameter}. 
When the initial state field $\lambda_i$ is close enough to the critical point $\lambda_i=1$, the first derivative of $\alpha$,  $d_{\lambda_i}\alpha$, exhibits a logarithmic divergence typical from the 2d-Ising universality class. 

In figure \ref{short_time}, we show the short time evolution of the Loschmidt echo for different quench protocols in both weak and strong regime. We see that it does not depend on the value of the final magnetic field $\lambda_f$ for times $t<t_{typ}$ as expected from \eqref{alpha}and observed in \cite{Muk12}. 

Figure \ref{parameter} shows the dependence of the Gaussian rate $\alpha$ as a function of $d$ for different quench protocols in the weak coupling case.

\subsection{Revival times}
\begin{figure}[h!]
\begin{center}
\includegraphics[scale=0.35]{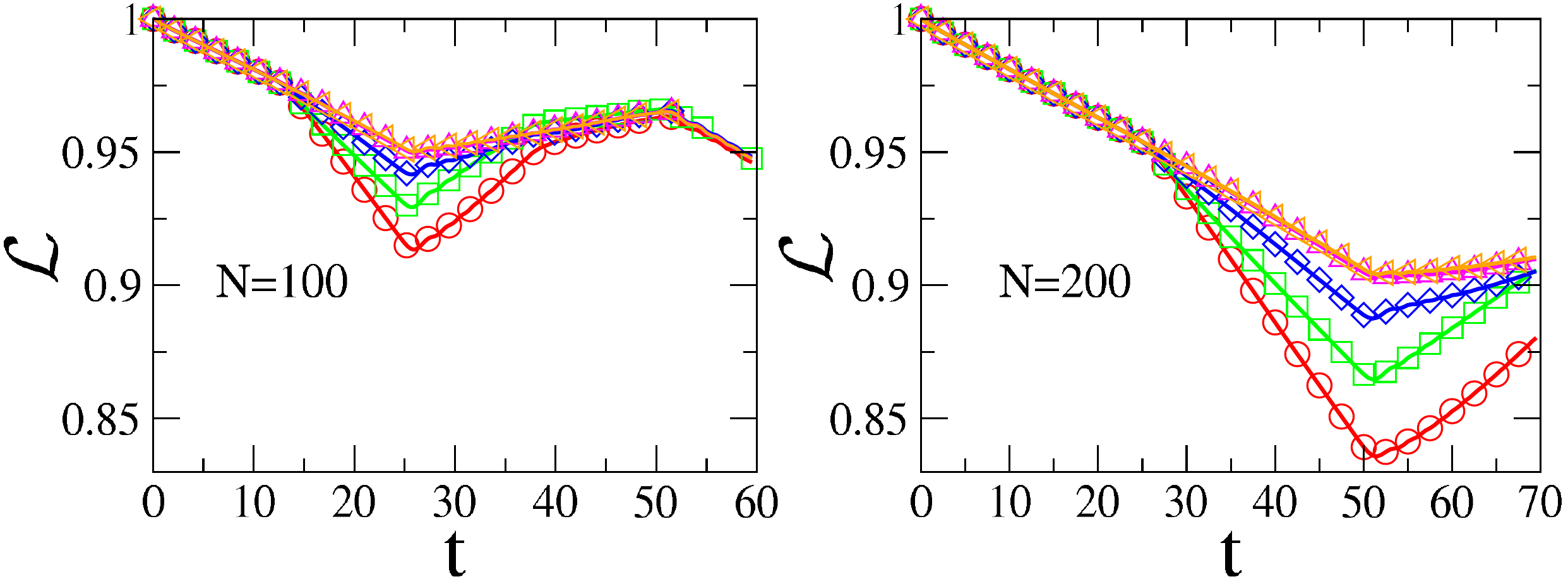}
\caption{(Color online) Loschmidt Echo for distances $d=N/2$ (red circles), $d=N/2-1$ (green squares), $d=N/2-2$ (blue diamonds), $d=N/2-5$ (magenta up triangles) and $d=N/2-15$ (orange left triangles) for $N=100$ (left) and $N=200$ (right). Note that due to their almost perfect matching, the two curves for $d=N/2-5$ and $d=N/2-15$ are not distinguishable. The other parameters are set to $\varepsilon=0.1$, $\lambda_i=1.5$ and $\lambda_f$=0.99. }
\label{revival}
\end{center}
\end{figure}

In the preceding section we have considered the short time behavior of the system, that is shorter than a revival time. However, 
depending on the separation distance $d$ and on the system size $N$ we observe a significant change of the Loschmidt echo for times of the order $N$.Note that the following considerations is exampled in the weak coupling case, but the same phenomenology of revival is observed in the opposite regime. For times $1\lesssim t<N/4$ when the initial state is not critical we observe a linear decay of the echo {whatever the final field is}. This is shown in figure \ref{revival} for systems of total size $N=100$ and $N=200$. We see in particular that when the separation distance of the two-qubits is far from the symmetric opposite position (that is $d=N/2$) the initial linear decay reverts to a linear increase at a revival time $t^*\simeq N/4$. The increase of the echo switches again to a linear decay after $t\simeq 2 t^*\simeq 2 \times N/4$, and so on.  

\begin{figure}[h!]
\begin{center}
\includegraphics[scale=0.34]{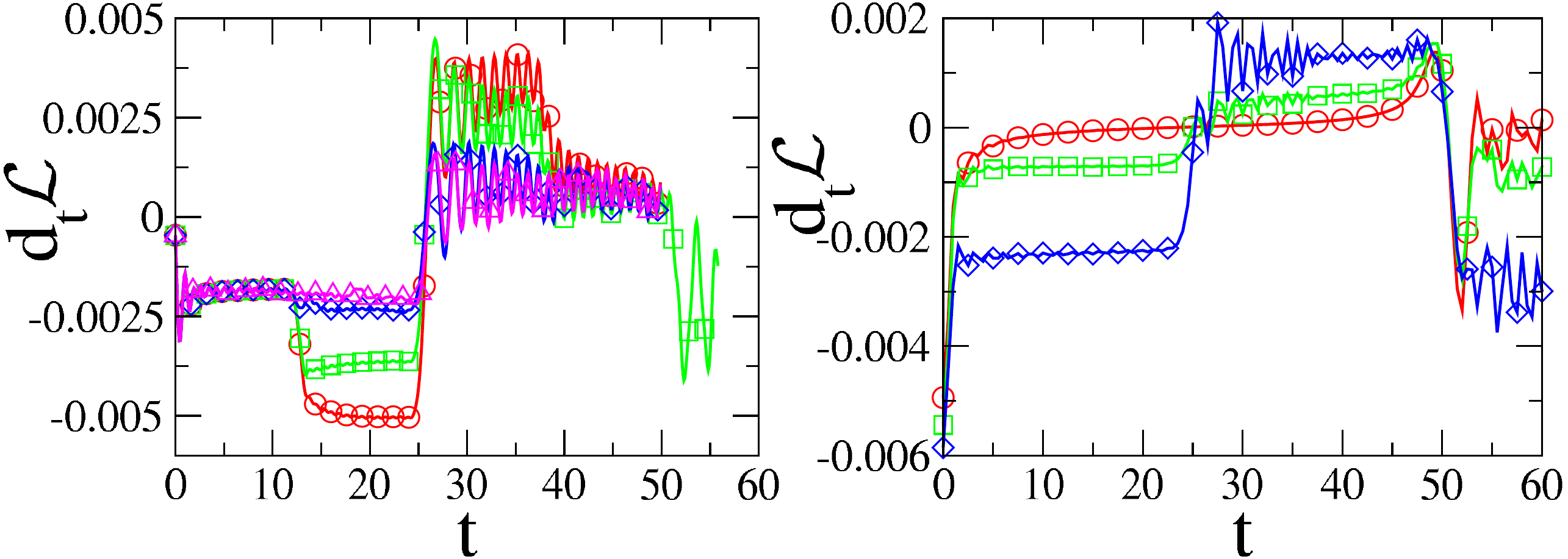}
\caption{(Color online) First derivative of the Loschmidt Echo with respect to time. In the left plot, we keep fixed $\lambda_i=1.5$ and $\lambda_f=0.99$ and we vary the distance. The different plots are $d=N/2$ (red circles), $d=N/2-1$ (green squares), $d=N/2-2$ (blue diamonds) and $d=N/2-15$ (magenta triangles) . In the right plot, the distance is $d=1$, $\lambda_f=0.99$ and $\lambda_i=0.99$ (red circles), $\lambda_i=0.9$ (green squares) and $\lambda_i=0.7$ (blue diamonds). The others parameters are $\varepsilon=0.1$ and $N=100$.}
\label{revival_der}
\end{center}
\end{figure}

When the separation distance $d$ comes close to the opposite location $N/2$, we observe a new singularity, emerging at half the original revival time, setting a new  time scale $\tau^*\simeq t^*/2\simeq N/8$.  This new  time scale $\tau^*$  is manifesting itself in a sudden speed-up of the linear decay until the revival time $t^*$ is reached.  The maximum slope of the new regime is reached  when the two qubits sit exactly on opposite sites along the chain, that is for $d=N/2$. This is best seen in the left panel of figure \ref{revival_der} which shows the numerical derivative of the echo for distances $d=N/2$, $N/2-1$, $N/2-2$, $N/2-5$ and $N/2-15$. One observes in particular that the new time scale $\tau^*$ has disappeared already for $d=N/2-5$ (see figure \ref{revival}). 
Note the remarkable feature that for whatever distance $d$ is,  at time $t=2t^*$ the Loschmidt echo recovers approximately the same value, as is clearly seen on the left panel of figure \ref{revival}.  

In the right panel of figure \ref{revival_der} we have plotted the evolution of the echo for two qubits at a distance $d=1$ for several quench protocols including the equilibrium situation $\lambda_i=\lambda_f$. We see that, contrary to the opposite location ($d=N/2$) situation there is no effect at $t=\tau^*$. One observes the revival phenomenon occurring at  $t^*\simeq N/4$ for the two non-equilibrium quenches considered here ($\lambda_i=0.7$, $0.9$ to $\lambda_f=0.99$). However, one clearly notice that in the equilibrium situation ($\lambda_i=\lambda_f$) the revival occurs at a time $t_{eq}^*$ which is twice the non-equilibrium revival time $t^*$.

\begin{figure}[!h]
\begin{center}
\includegraphics[scale=0.5]{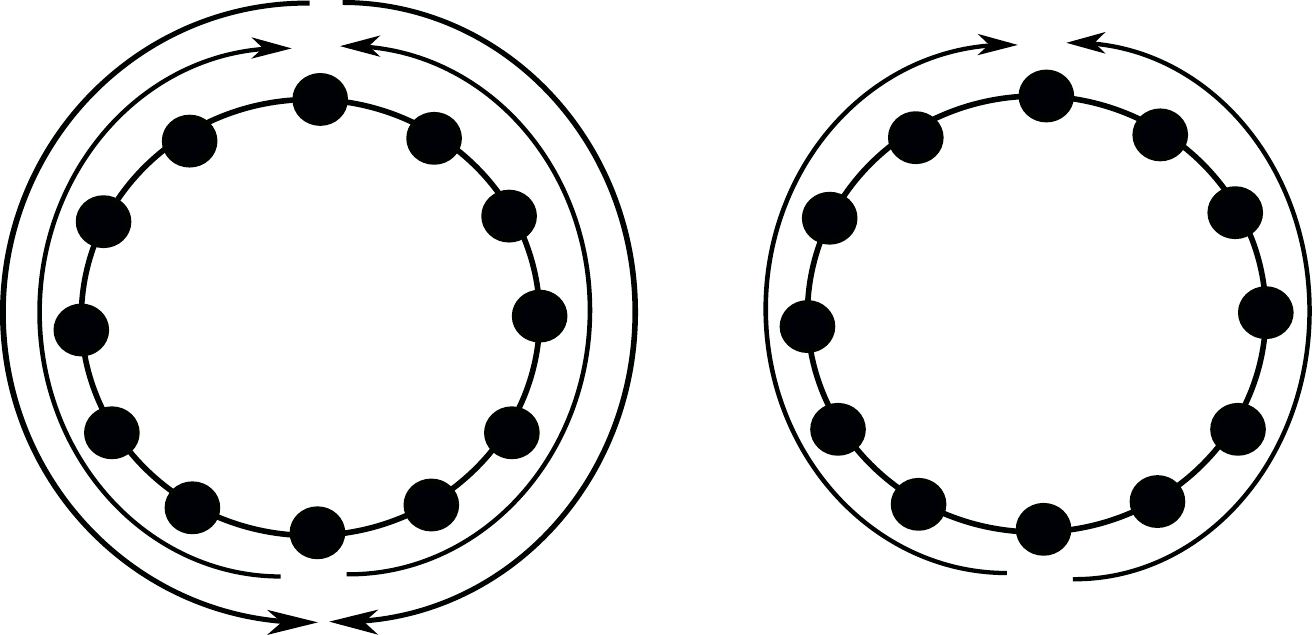}
\caption{Pictural representation of the difference between the global (left) and the local quench (equilibrium, right plot). In the quenched case, the excitations are emitted from everywhere, in particular in one spin and its opposite. Then, the revival time is the time required for the excitations to travel on a distance which is the half of the chain. On the contrary, in the equilibrium situation, the excitations are emitted only in one position, and the revival time is the time needed to travel along the entire chain.}
\label{revival_picture}
\end{center}
\end{figure}

The fact that in the non-equilibrium quench case ($\lambda_i\ne\lambda_f$) the revival time is twice shorter than in the equilibrium situation ($\lambda_i=\lambda_f$) can be understood in the following way \cite{cal05}: Indeed, the non-equilibrium situation corresponds to a global quench. At each position of the chain the energy is suddenly changed and from every point pairs of free quasi-particles are emitted with opposite momenta $\pm k$. The fastest particles travel with velocities 
\begin{equation}
v_{g}=\max_k\left.\left(\frac{\partial\varepsilon_k}{\partial k}\right)\right|_k=
\left \{
\begin{array}{c }
    2\lambda_f \quad \textrm{if} \quad \lambda_f<1 \\
    2 \quad \textrm{if} \quad \lambda_f\ge1 \\
\end{array}
\right.,
\label{group_velocity}
\end{equation}
and since all chain sites behave as local emitters after a time $t^*=\frac{1}{2} N/v_g$ the configuration of quasiparticles along the chain is starting to restore its initial state, leading to the increase of the echo.
On the contrary the equilibrium case corresponds to a local quench at the qubits positions. In that case, quasi-particles are emitted only on that localized sites and they need to circle at least once along the full chain to reconstruct the initial state, such that $t^*=N/v_g$. These quasi-particle interpretation is depicted schematically on figure  \ref{revival_picture}. 

\begin{figure}[!h]
\begin{center}
\includegraphics[width=1\linewidth]{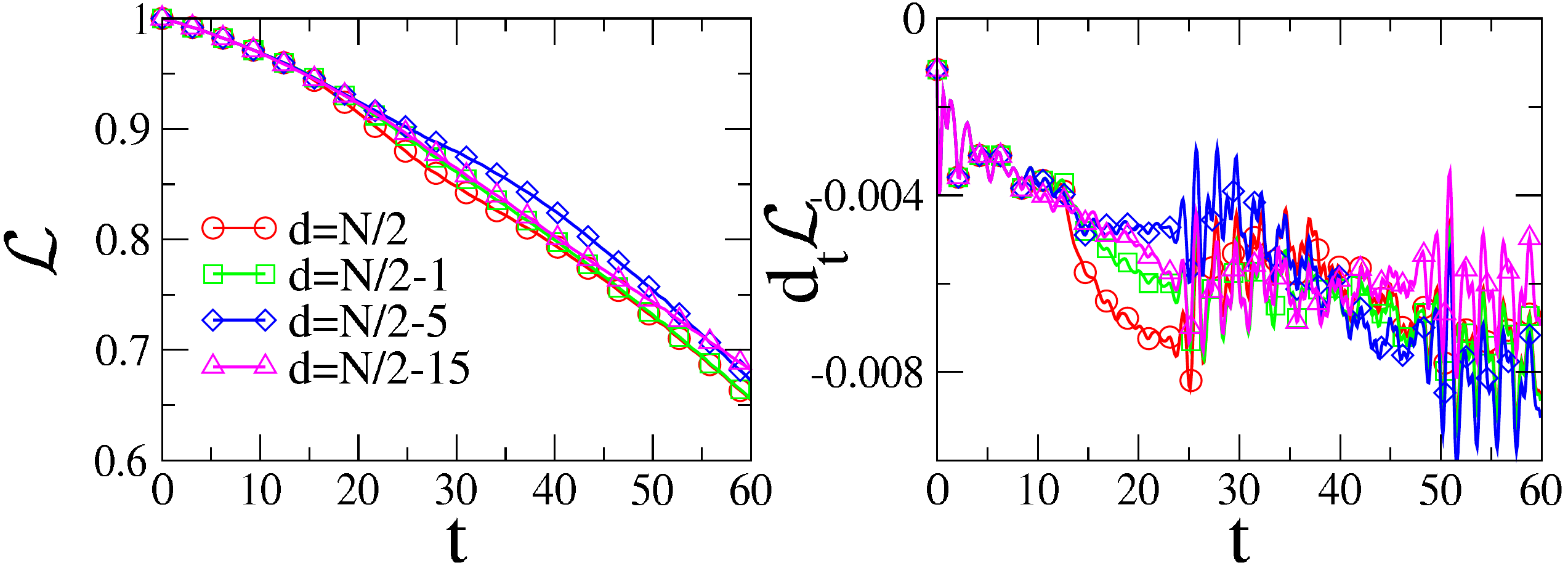}
\caption{ (Color online) Left: Loschmidt echo for a critical initial environment for distances $d=N/2$ (red circles), $d=N/2-1$ (green squares), $d=N/2-5$ (blue diamonds) and $d=N/2-15$ (magenta triangles). Right: Time derivative of the Loschmidt echo for the previous distances. Other parameters are $N=100$, $\lambda_f=1.5$ and $\varepsilon=0.1$.}
\label{revival_crit}
\end{center}
\end{figure}

When the starting state is long-range, that is for an initial field value $\lambda_i$ very close to the critical value $\lambda_c=1$, the revival phenomenology is very similar to what has already been discussed: At symmetric positions of the defect qubits ($d\simeq N/2$),  one observes a singular behavior of the echo at time $\tau^*=t^*/2$ and a revival phenomenon starting at $t^*$. Far from the symmetric position, the singular behavior at $\tau^*$ has disappeared  and just the revival time $t^*$ shows up. For the non-equilibrium quench ($\lambda_i\neq \lambda_f$) the revival time $t^*=N/4$ while for the equilibrium case ($\lambda_i=\lambda_f$) the revival time $t^*=N/2$ is twice bigger. The main difference to the non-critical initial state lies in the fact that the shape of the decay  (and increase) of the Loschmidt echo  is no longer linear as it was for an initial short-range state, see figure \ref{revival_crit}.

\subsection{Comparison to the independent dynamics}
\begin{figure}[!h]
\begin{center}
\includegraphics[scale=0.33]{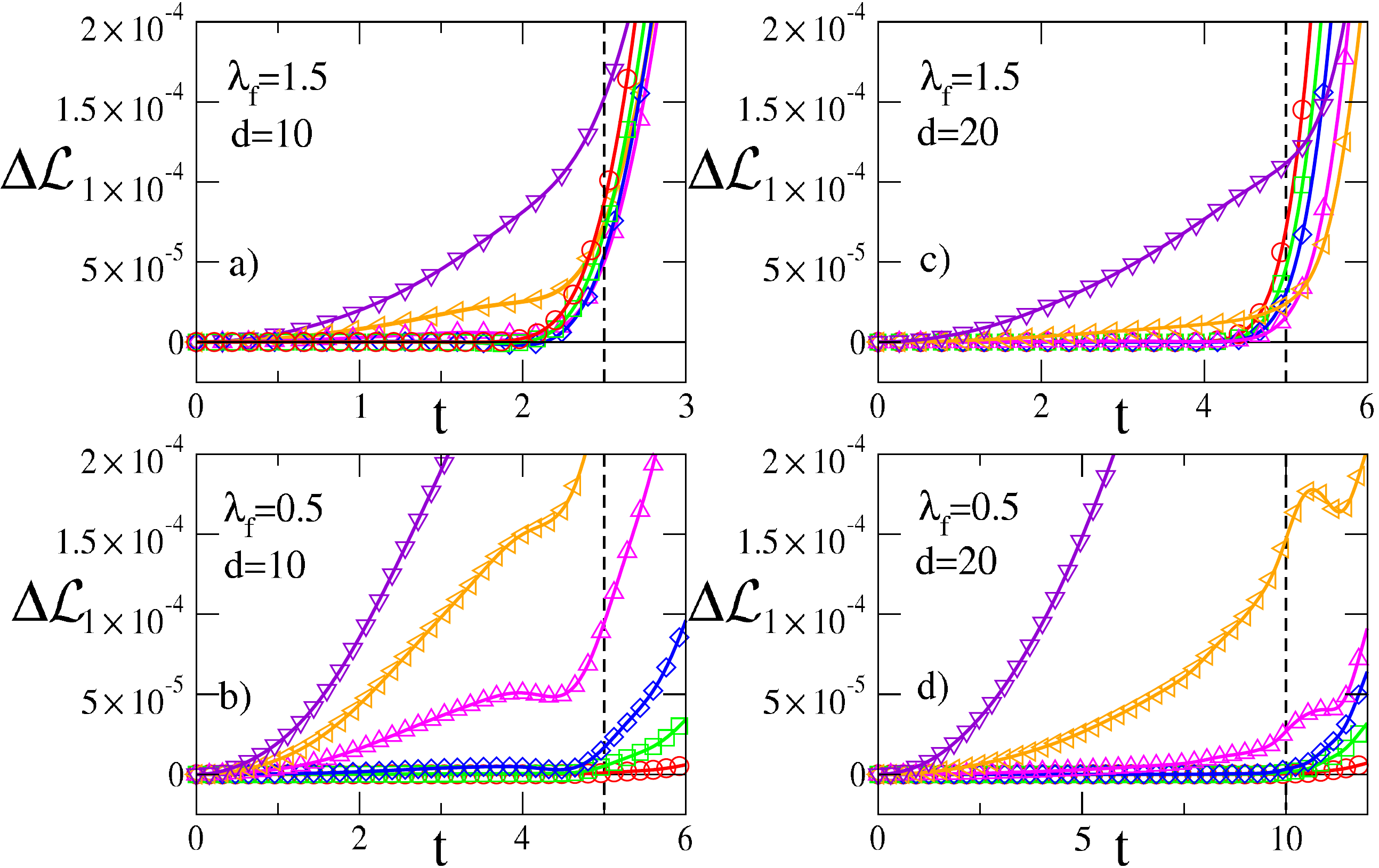}
\caption{(Color online) Difference between the Loschmidt echo $\Delta \mathcal{L}$ in the situation where the two spins are coupled to the same bath and to two independent baths as a function of time for different quench protocols and distances. The initial magnetic fields are: $\lambda_i=0.4$ (red circles in b) and d)), $\lambda_i=0.5$ (red circles in a) and c)), $\lambda_i=0.7$ (green squares), $\lambda_i=0.8$ (blue diamonds), $\lambda_i=0.9$ (magenta up triangles), $\lambda_i=0.95$ (orange left triangles) and $\lambda_i=1$ (indigo down triangles). In all plots, we also add in dashed line the theoretical value of $t_{ind}=d/{(2v_g)}$}
\label{ind}
\end{center}
\end{figure}

Part of the disentanglement observed between the two qubits is a consequence of their direct coupling to the environment and the other part comes from their mutual interaction, mediated through the bath degrees of freedom. In order to quantify the part of the decoherence that comes from this direct coupling  we compute the difference of the Loschmidt echo between the situation where the spins are coupled to a common environment and the limiting case of two spins coupled to two independent ones: $\Delta\mathcal{L}=\mathcal{L}-\mathcal{L}_{ind}$. The results are presented in figure \ref{ind} where we have plotted $\Delta \mathcal{L}$ as a function of time for different quench protocols and distances $d$.

For initial magnetic fields far enough from the critical field, the difference $\Delta\mathcal{L}$ is equal to zero up to a time $t_{ind}$ after which $\mathcal{L}$ and $\mathcal{L}_{ind}$ starts to differ significantly. This implies that for times shorter than $t_{ind}$, the two spins are evolving independently like if they were coupled to non-interacting bath. After $t_{ind}$, the two spins start to interact through the chain and their evolution is no longer independent. Note that this time is not dependent on the initial magnetic fields, but rather depends on the final one and of course on the distance between the two defect spins. This can be understood in the following way: the two spins will evolve independently until an entangled pair of excitations created by the quench in the middle of the two qubits  has reached them and consequently correlating them. The time required for this pair of excitation to travel along the chain is given by $t_{ind}= (d/2)/ v_g$ where the velocity $v_g$ is given by \eqref{group_velocity} and depends only on $\lambda_f$. Notice that in the equilibrium situation, the fact that the quasi excitations are emitted at positions $0$ and $d$ leads to a $t_{ind}$ twice bigger. The time $t_{ind}$ is indicated in figure \ref{ind} by the vertical dashed lines. We see that this prediction is in a quite good agreement with the numerical data. 

On the other hand, when the initial magnetic field is close to the critical value $\lambda_i=1$, there is already a non vanishing difference $\Delta\mathcal{L}$ at $t=0^+$ due to the long-range correlations present in the chain. The typical correlation length in the Ising chain is given by $\xi=|\ln(\lambda_i)|^{-1}$ \cite{pf70} and if the distance $d$ separating the two defect qubits is smaller than this correlation length $\xi$, the two defects are no longer independent already at $t=0$. This is clearly seen in figure \ref{ind} for $\lambda_i= 0.95$ and $1$ where we see the large departure of $\Delta  \mathcal{L}$ from 0. Moreover, at a fixed initial field $\lambda_i$ (that is at a fixed correlation length $\xi$), the larger the separation distance $d$ between the two defects spins, the smaller the departure from 0 of $\Delta  \mathcal{L}$   as seen by comparing the left panels of figure \ref{ind}, where the distance was fixed to $d=10$ to the right panels $d=20$ in the left one.  
Nevertheless, the signature of the correlation of the qubits through the entangled pair emission mechanism, discussed above for short range initial states, is also present in this critical case. We observe clearly on figure \ref{ind} a significant deviation of $ \mathcal{L}$ to $\mathcal{L}_{ind}$ for times larger than $t_{ind}$.

\section{Conclusion and Summary} 
We have investigated the effect on the disentanglement of two qubits initially prepared in a Bell state of a global quench of an Ising chain environment to which the qubits are coupled.  
We have in particular studied the dependance of the decoherence on the distance separating the two qubits. 
We have shown that the decoherence of the qubits is enhanced at large times in the quenched environment case with respect to the equilibrium chain considered in \cite{Cor08}. 
We have seen that the bigger the quench amplitude is the stronger the decoherence is, such that the quenched situation leads always to an increased qubits decoherence. When the initial state of the Ising chain environment is close to criticality  
the Loschmidt echo exhibits a clear signature of the long range nature of the initial state. At long times, of order of the environment size (the number of sites $N$ of the ITF),  we observe a revival phenomenology in the Loschmidt echo starting at a time $t^*$ which is twice shorter than that of the equilibrium case. This is explained in terms of the propagation of quasi-particles emitted, due to the global quench, at every sites of the ITF chain, contrary to the equilibrium situation where only the sites directly coupled to the two qubits act as quasi-particles emitters. As a consequence of the propagation of the quasi-particles in the chain, they have to travel half the chain length in order to rebuild the initial correlations while they have to circle around the full chain in order to start to rebuild correlations in the equilibrium case. Finally, one observe an intriguing phenomenon when the qubits are coupled on opposite sites of the ITF chain, that is when they are maximally separated, indeed there is singular behavior appearing in the Loschmidt echo at half the revival time scale, $t^*$, which does not seem to be explainable in terms of the quasi-particles propagation but is rather an interference effect.

\section*{Acknowledgements}
We are grateful to Giovanna Morigi and Cecilia Cormick for helpful discussions. 
P.W benefited from the support of the International Graduate College on Statistical Physics of Complex Systems between the universities of Lorraine, Leipzig, Coventry and Lviv.

\appendix


\bibliographystyle{apsrev}
\bibliography{biblio}

\end{document}